\documentclass[prl,twocolumn,aps,amsmath,amssymb,showpacs]{revtex4}
\usepackage{graphicx}
\usepackage{dcolumn}
\usepackage{bm}
\usepackage{wrapfig}
\usepackage{amsmath,amsfonts,amssymb,bm,ulem}
\usepackage[usenames]{color}
\begin{document}

\title{Asymmetric Berry-Phase Interference Patterns in a Single-Molecule Magnet}

\author{H. M. Quddusi$^1$}
\author{J. Liu$^2$}
\author{S. Singh$^1$}
\author{K. J. Heroux$^4$}
\author{E. del Barco$^1$}
\author{S. Hill$^3$}
\author{D. N. Hendrickson$^4$}

\affiliation{$^1$Department of Physics, University of Central Florida,
Orlando, FL 32816, USA}
\affiliation{$^2$Department of Physics, University of Florida, Gainesville, FL 32611, USA}
\affiliation{$^3$National High Magnetic Field Laboratory and Department
of Physics, Florida State University, Tallahassee, FL 32310, USA}
\affiliation{$^4$Department of Chemistry and Biochemistry, University of
California at San Diego, La Jolla, CA 92093, USA}

\begin{abstract}
A Mn$_4$ single-molecule magnet displays asymmetric Berry-phase interference patterns in the transverse-field ($H_T$) dependence of the magnetization tunneling probability when a longitudinal field ($H_L$) is present, contrary to symmetric patterns observed for $H_L=0$. Reversal of $H_L$ results in a reflection of the transverse-field asymmetry about $H_T=0$, as expected on the basis of the time-reversal invariance of the spin-orbit Hamiltonian which is responsible for the tunneling oscillations. A fascinating motion of Berry-phase minima within the transverse field magnitude-direction phase space results from a competition between non-collinear magneto-anisotropy tensors at the two distinct Mn sites.
\vspace{-2 mm}
\end{abstract}

\maketitle

{\noindent Almost} two decades of research have established single-molecule magnets (SMMs) as prototype systems for understanding fundamental quantum  phenomena associated with nanoscale magnetism \cite{chudnovsky98,villain08}, as well as demonstrating their potential for future applications \cite{leuenberger01}. The most important characteristics of SMMs can be modeled reliably using a giant-spin approximation (GSA) whereby the molecule is treated as a rigid magnetic unit with total spin, $\it{S}$, weakly interacting with its enviroment. Indeed, this model accounts for the essential features of the quantum tunneling of magnetization (QTM) observed in these molecules, as well as its quenching due to Berry phase interference (BPI) resulting from different tunneling trajectories \cite{WWSess99,delbarco03,foss09}. However, the GSA ignores the internal couplings within a SMM, thereby completely failing to account for QTM transitions that involve fluctuations of the total spin of the molecule \cite{ramsey08,carreta08,bahr08}, or otherwise obscuring intrinsic relationships that exist between QTM selection rules and the underlying molecular structure \cite{henderson09}. In a recent interesting example, it was demonstrated that a tilting of the zero-field splitting (zfs) tensors in a triangular Mn$_3^{\rm III}$ SMM (lowering the symmetry of the spin Hamiltonian from $C_{6}$ to $C_3$) results in new QTM selection rules and strongly affects the transverse field dependence of the remaining forbidden QTM resonances \cite{henderson09}. These observations likely explain the absence of QTM selection rules in most SMMs studied to date, since internal dipolar fields and/or weak sample disorder are often sufficient to cause observable relaxation at resonances otherwise forbidden by symmetry. A detailed understanding of these and related phenomena has mostly been facilitated by studies of low nuclearity SMMs \cite{hill10,Wilson}, where exact diagonalization of the multi-spin Hamiltonian enables consideration of the internal degrees of freedom of the molecule. 

\begin{figure}
\includegraphics[width=8.6cm]{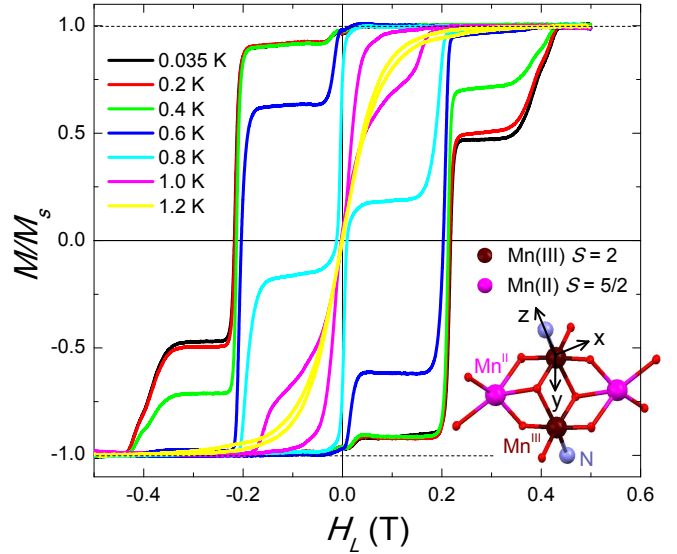}
\vspace{-6 mm} \caption{(Color on-line)  Hysteresis loops recorded as a function of $H_L$ at different temperatures. The inset shows the relevant magnetic centers (i.e. Mn$^{\rm II}$, Mn$^{\rm III}$, N and O) at the core of the Mn$_4$-Bet unit. The principal magnetic axes of the molecule are indicated, with the easy ($z$-) axis determined by the JT elongation associated with the Mn$^{\rm III}$ ions (Mn$^{\rm III}$-N line, see main text); the $y$-axis lies in the plane defined by the two Mn$^{\rm III}$ ions and their JT axes.}
\vspace{-4 mm}
\end{figure}

In this letter, we focus on the QTM relaxation associated with a centro-symmetric mixed-valent Mn$_2^{\rm II}$Mn$_2^{\rm III}$ complex which shows an asymmetric BPI pattern with respect to the polarity of the transverse component of the applied field ($H_T \perp$ magnetic easy axis). We show that this behavior results from a competition between non-collinear magneto-anisotropy tensors at the two crystallographically distinct Mn ions, which is also responsible for an unusual motion of the Berry-phase minima within the transverse field magnitude-direction phase space. We show how the asymmetry can be inverted upon reversal of the longitudinal field ($H_L \parallel$ easy axis), i.e., the BPI pattern is invariant with respect to a full inversion of the applied field, consistent with the time-reversal symmetry of the underlying zero-field Hamiltonian. 

The [Mn$_4$(Bet)$_4$(mdea)$_2$(Hmdea)$_2$](BPh$_4$)$_4$ complex (henceforth Mn$_4$-Bet) crystallizes in the triclinic $P\bar 1$ space group with half the molecule in the asymmetric unit; the other half is generated via inversion, resulting in the four Mn ions lying in a plane (the molecular plane), with the Mn$^{\rm III}$ Jahn-Teller (JT) axes oriented along the Mn$^{\rm III}$-N bonds, which lie 122.61 degrees out of this plane, i.e., roughly perpendicular to the molecular plane \cite{database,heroux10}. There are no solvent groups in the lattice and the four BPh$_4^-$ anions enhance isolation, resulting in extremely clean X-ray diffraction and EPR data \cite{heroux10,EPAPS}. A sketch of the Mn$_4$ core, where the magnetic axes are indicated, is inset to Fig. 1. Magnetic and EPR measurements performed at relatively high temperatures ($T>2$~K) suggest a spin $S=9$ ground state, and that Mn$_4$-Bet is a SMM with a barrier of $\sim20$~K \cite{heroux10}.

A high-sensititivity micro-Hall effect magnetometer, a He$^3$/He$^4$ dilution fridge and a 3D vector superconducting magnet were employed to record magnetization hysteresis curves as a function of a magnetic field applied parallel to the easy axis of the molecules \cite{note0}, at temperatures down to 35~mK. The results are shown in Fig.~1, where extremely sharp QTM resonances ($k=0$, 1 $\&$ 2), spaced by $\Delta H_L \approx 0.21$~T, confirm the high quality of the crystal. Within the GSA, this spacing corresponds to an axial zfs parameter, $D=-0.28$~K ($g=2$). The observed blocking and crossover temperatures are $\sim$1.2~K and $\sim$0.2~K, respectively. A transverse field was subsequently employed in order to study the symmetry of the QTM in resonances $k=0$ and $k=1$. Fig.~2a shows the modulation of the QTM probability, $P_{k}=(M_f-M_i)/(M_{sat}-M_i)$ \cite{note1}, for resonance $k=0$, as a function of $H_T$ applied along the magnetic hard axis ($\phi=0^\circ$). This angle, which lies $\sim30^\circ$ away from one of the crystal faces, was deduced from the two-fold modulation of $P_{0}$ as a function of the orientation, $\phi$, of a 0.2~T transverse field within the hard plane (see inset to Fig.~2a)~\cite{noteA}.

The $P_0$ oscillations in Fig.~2a correspond to BPI, with minima at regularly spaced field values ($\Delta H_T=0.3$T). A maximum in $P_0$ is found at $H_T=0$, as expected for an integer spin value. Within the GSA, $\Delta H_T=2k_B(2E[E+D])^{1/2}/g\mu_B$  \cite{garg93}, yielding a 2$^{\rm nd}$-order rhombic zfs parameter, $E=\pm 60$~mK. Note that the regularly spaced $k=0$ BPI minima are invariant under inversion of $H_T$, i.e., they are symmetric with respect to $H_T=0$. Interestingly, this is not the case in resonance $k=1$, for which the behavior of the QTM probability is very different. This can be seen in Fig.~2b, which illustrates the dependence of $P_1$ on $H_T$, for $\phi=13.5^\circ$ (the angle for which the first BPI minimum at $H_T=0.30$~T is the sharpest). In fact, for resonance $k=1$, different BPI minima appear at different field orientations, $\phi$, of the transverse field within the $xy$ (hard) plane of the molecule (see Fig.~3)~\cite{noteA}, i.e., the first minimum ($H_T=0.3$~T) appears at $\phi=13.5^\circ$, while the second ($H_T=0.6$~T) occurs at $\phi=6^\circ$, contrary to what is found for the $k=0$ resonance (all $P_0$ minima are seen most clearly at $\phi=0^\circ$). Such behavior has been predicted theoretically~\cite{JLTP,Miyashita}, though never observed experimentally.

\begin{figure}
\includegraphics[width=8.6cm]{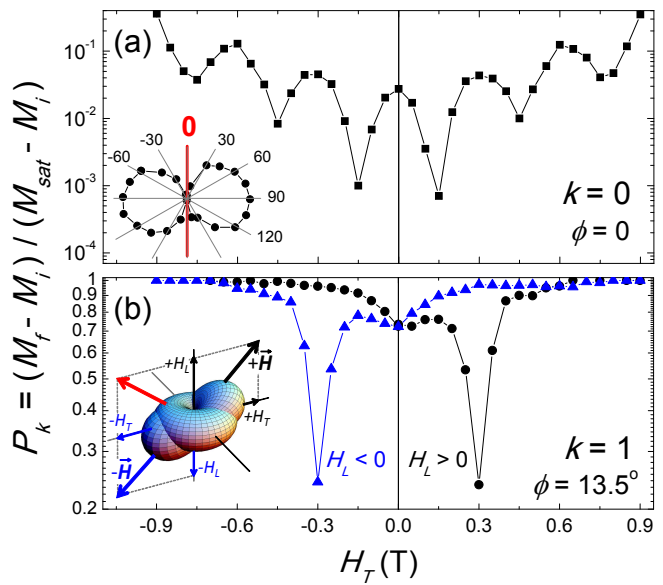}
\vspace{-6 mm} \caption{(Color on-line)  Modulation of the QTM probabilities for resonances $k=0$ (a) and $k=1$ (b) as a function of $H_T$ applied at different angles, $\phi$, within the hard plane of the Mn$_4$-Bet SMM~\cite{noteA}. The inset to (a) illustrates the two-fold angular modulation of $P_0$ for $H_T=0.2$~T, providing clear evidence for a significant $2^{\rm nd}$-order rhombic anisotropy. The asymmetry of the BPI pattern of oscillations in resonance $k=1$ is inverted upon reversal of $H_L$. The inset to (b) illustrates the classical anisotropy barrier generated by the non-collinear zfs tensors (see main text for explanation) and the different perspectives resulting from permutations of $\pm H_T$ and $\pm H_L$.}
\vspace{-2 mm}
\end{figure}

Before considering this aspect in detail, we first discuss the asymmetric nature of the BPI oscillation pattern in resonance $k=1$. As seen clearly in Fig.~2b, reversal of the longitudinal field, $H_L$, results in a reflection of the $P_1$ BPI pattern about $H_T=0$. In other words, the BPI minima {\it are} in fact invariant under a full magnetic field inversion, as required on the basis of the time-reversal invariance of the spin-orbit Hamiltonian responsible for this physics. As noted above, the symmetries of BPI patterns must respect the symmetry of the zero-field spin Hamiltonian. If one considers only 2$^{\rm nd}$-order zfs within the GSA, then the resulting Hamiltonian necessarily belongs to the orthorhombic point group and possesses the following symmetry elements: (1) three mutually orthogonal two-fold rotation axes ($x$, $y~\&~z$); (2) three mutually orthogonal mirror planes ($xy$, $xz~\&~yz$); and (3) an inversion center. (2) guarantees invariance with respect to reversal of $H_T$, i.e., it enforces symmetric BPI patterns, irrespective of whether a longitudinal field is applied ($k>$~0) or not ($k=0$). As we show below, one must break the $xy$ mirror symmetry in order to obtain asymmetric BPI patterns with respect to inversion of $H_T$. In this case, reversal of $H_L$ results in different patterns; the time reversal symmetry then guarantees that these two patterns are mirror images. Nevertheless, no matter how many spatial symmetries are broken, the time-reversal invariance of the spin-orbit interaction guarantees that the BPI minima should be invariant under a full reversal of the applied field, i.e., simultaneous reversal $H_L$ and $H_T$, as we have confirmed experimentally. 

It is possible to reproduce the essential features of the experiments by introducing $4^{\rm th}$-order terms into the GSA; the $xy$ mirror symmetry can then be broken by rotating the coordinate frames of the $2^{\rm nd}$ and $4^{\rm th}$-order tensors. Interestingly, this approach also reproduces the complex motion of the $P_1$ minima within the $H_T-\phi$ phase-space shown in Fig~3. However, a complete GSA analysis for Mn$_4$-Bet requires many parameters and provides little insight, while the same physics can be naturally understood within a multispin description which considers the internal structure of the Mn$_4$-Bet molecule. Note that the emergence of significant higher-order anisotropy terms within the GSA is a manifestation of mixing of the ground spin state with excited states, which can only be captured within a multi-spin model~\cite{hill10,Wilson}. In this context, the $xy$ mirror symmetry may be trivially broken by rotating (tilting) the zfs tensors at the two inequivalent magnetic sites in the molecule so that their local $z$-axes no longer coincide (this is similar to the case of the Mn$^{\rm III}_3$ SMM discussed in the introduction~\cite{henderson09,Junjie}). 

To explain the experimental findings we have diagonalized the multi-spin Hamiltonian, where the four Mn ions are coupled according to the sketch in Fig.~4a.
\begin{align}
H=\sum_{i = 1}^{3}& (\vec{s}_i\cdot\tilde{R}^T_i\cdot\tilde{d}_i\cdot\tilde{R}_i\cdot\vec{s}_i- g\mu_B \vec{s}_i\cdot\vec{B})
\nonumber\\
+\sum_{i,j (i>j)} &(\vec{s}_i\cdot\tilde{J}_{i,j}\cdot\vec{s}_j
-\frac{\mu_0(g\mu_B)^2}{4\pi r_{i,j}^3}\vec{s}_i\cdot\tilde{\Delta}\cdot\vec{s}_j).
\label{Eqn1}
\end{align}

{\noindent The first} term represents the local anisotropy of the $i^{\rm th}$ ion, $\tilde{d}_i$ being the $2^{\rm nd}$-order zfs tensor ($d_i,_{xx}=e_i$, $d_i,_{yy}=-e_i$ $\&$ $d_i,_{zz}=d_i$, with $d_i$ $\&$ $e_i$ representing the axial and rhombic anisotropies, respectively); the $\tilde{R}_i$ are Euler matrices (defined by Euler rotation angles $\alpha_i$, $\beta_i$ $\&$ $\gamma_i$) specifying the orientations of these tensors. The second term is the Zeeman coupling to the applied field, where we assume an isotropic Land{\'e} factor, $g=2.00$. The third and fourth terms represent the exchange and dipolar interactions, respectively. These terms are also time-reversal invariant, and do not change any of the preceeding arguments.  As depicted in Fig.~4a, three independent near-neighbor exchange coupling constants, $J_a$, $J_b$ and $J_c$, are considered: $J_{1,2}=J_{3,4}=J_a$, $J_{2,3}=J_{1,4}=J_b$, $J_{2,4}=J_c$ and $J_{1,3}=0$. The dipolar matrix, $\tilde{\Delta}_{i,j}$ has been chosen to exactly reproduce all of the dipolar couplings within the molecule, with no fitting parameters.

\begin{figure}
\includegraphics[width=8.6cm]{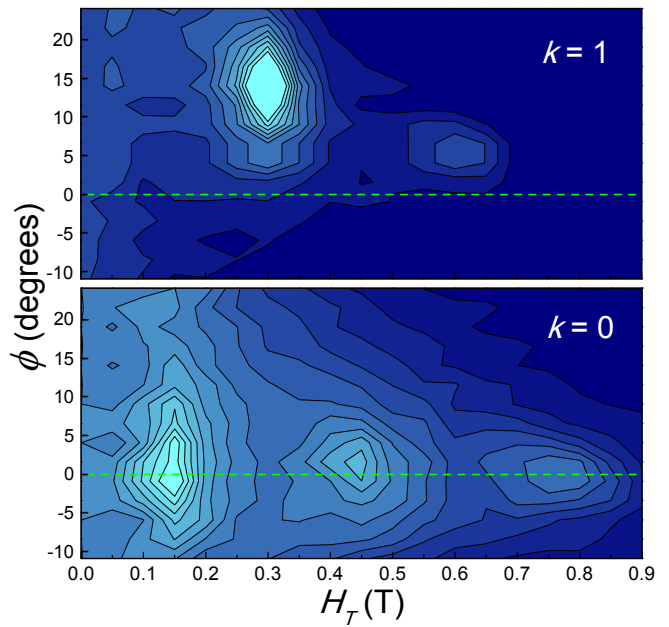}
\vspace{-6 mm} \caption{(Color on-line)  Contour plots of the QTM probabilities for resonances $k=0$ and $k=1$ as a function of $H_T$ and $\phi$. All of the $k=0$ minima lie approximately along the $\phi=0^\circ$ axis, whereas the $k=1$ minima appear at different orientations for different $H_T$ values.}
\vspace{-2 mm}
\end{figure}

Figure 4b plots the locations of BPI minima obtained via diagonalization of Eqn.~\ref{Eqn1} (solid red symbols). This simulation, which takes into account the small misalignement of the experimental field rotation plane~\cite{noteA}, employed the following parameters: $d_2=d_4=-4.99$~K and $e_2=e_4=0.82$~K, with the easy and hard anisotropy axes along $z$ ($\alpha_2=0$) and $x$ ($\beta_2=0$), respectively; $d_1=d_3=-0.67$~K $\&$ $e_1=e_3=0$, with the axes rotated with respect to the central spin by identical Euler angles $\alpha_{1,3}=45^\circ$, $\beta_{1,3}=0^\circ$ (as required by inversion symmetry); $\gamma$ being zero for all ions; finally, isotropic ferromagnetic exchange constants $J_a=-3.84$~K, $J_b=-1.20$~K and $J_c=-3.36$~K are used. It should be stressed that these parameters are additionally constrained by the locations of hysteresis loop steps (Fig.~1) and extensive angle-dependent EPR measurements~\cite{EPAPS}. Moreover, the obtained anisotropy values for the Mn$^{\rm III}$ ions are very similar to related Mn$^{\rm III}$ complexes~\cite{henderson09}, while the $d_{1,3}$ value lies within the bounds reported for other Mn$^{\rm II}$ systems~\cite{MnIIref}. The quantitative agreement with experiment is also excellent. The motion of the $P_1$ minima can be understood as a result of the competition between different anisotropic interactions within the molecule, without a need to invoke unphysical $4^{\rm th}$ and higher order anisotropies. Importantly, the angular positions ($\phi$) of the $k=1$ minima move with $H_T$, while the $k=0$ minima remain stationary, as found experimentally (Fig.~3).  

\begin{figure}
\includegraphics[width=8.6cm]{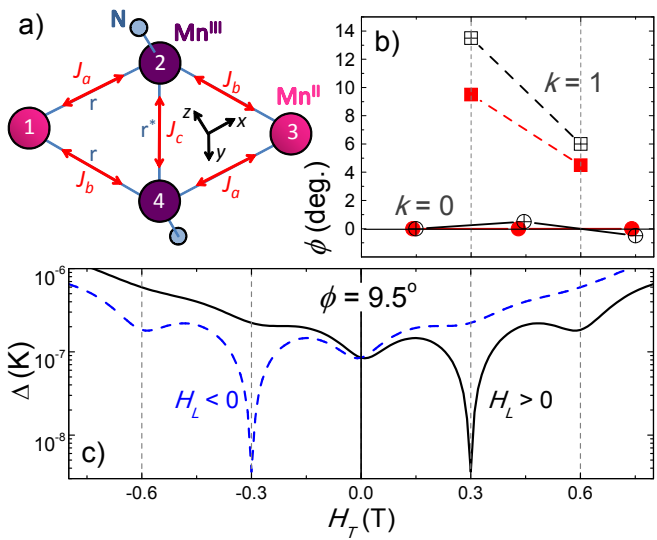}
\vspace{-6 mm} \caption{(Color on-line)  (a) Sketch of the Mn$_4$ core indicating the interaction parameters used to explain the results. (b) Measured (open black symbols) and calculated (solid red symbols) $\phi/H_T$-dependence of the BPI minima for resonances $k=0$ (circles) and 1 (squares) obtained from Eqn.~\ref{Eqn1}. (c) Calculated tunnel splittings for resonance $k=1$, for $\phi=9.5^\circ$, as a function of $H_T$ for $H_L>0$ (continuous balck line) and $H_L<0$ (discontinuous blue line).}
\vspace{-2 mm}
\end{figure}

Finally, the multi-spin model perfectly reproduces the $H_T$ asymmetry of the $k=1$ BPI pattern. As seen in Fig.~4c, the asymmetry is reversed upon inversion of $H_L$, as required by the time-reversal invariance of the anisotropic interactions in Eqn.~\ref{Eqn1}, and observed experimentally (Fig. 2b). The crucial ingredient is the tilting of the zfs tensors of the external spins, $s_1$ $\&$ $s_3$, relative to the central spins $s_2$ $\&$ $s_4$, so that the $xy$ mirror symmetry is broken.  This is illustrated in the inset to Fig.~2b, where one observes that the classical energy landscape is invariant under full field inversion (blue vs. black arrows), while this is not the case when only $H_T$ is reversed (red vs. black arrows). The Euler angle $\alpha_{1,3}=45^\circ$ results in a significant projection of the relatively weak anisotropy associated with the Mn$^{\rm II}$ ions into the hard ($xy$) plane. This, together with the finite $e_{2,4}$ parameters and the dipolar interactions, results in competing transverse interactions and to the complexity of the BPI patterns observed in Fig.~3. We note that the dipolar interaction has a very significant effect on the energy levels of the molecule: the zfs within the $S = 9$ multiplet varies by as much as $10\%$ when dipolar interactions are omitted, and the location of the $k = 1$ QTM step is shifted by $\sim 0.02$T.

We conclude by noting that asymmetric BPI patterns have been seen in other centro-symmetric SMMs for which a clear explanation has been lacking~\cite{ramsey08,bahr08,delbarco10}. The present results may help shed light on the effect that symmetric anisotropic interactions can have in magnetic systems with inversion symmetry, where a net antisymmetric interaction is strickly forbidden. The present work clearly demonstrates how studies of simple low nuclearity systems can address fundamental symmetry considerations related to QTM in molecular nanomagnetism.

\section{Acknowledgements}
\label{sec:end}
The authors acknowledge support from NSF, specifically H.M.Q, S.S and E.d.B. from DMR-0747587, J.L. and S.H from DMR-0804408, and K.J.H and D.N.H. from CHE-0714488, and from the University of Florida High-Performance Computing Center for providing computational resources and support (URL: http://hpc.ufl.edu).


\begin{thebibliography}{99}
\bibitem{note0} The magnetic axes were determined experimentally from the angular field behavior of the QTM resonances.
\bibitem{note1} $M_i$ and $M_f$ are the magnetizations before and after the resonance, respectively. $M_{sat}$ is the saturation value.
\bibitem{noteA} We note that there was a small ($\sim 3^\circ$) misalignment of the field rotation plane during the measurements. This slightly influences, but is not the primary cause of, the BPI minima motion. Simulations demonstrate that similar results are obtained when $H_T$ is rotated exactly within the $xy$ plane, for which the $k=1$ first and second BPI minima are found at $\phi=7.5^\circ$ and $\phi=3^\circ$, respectively.
\end{thebibliography}
\end{document}